\documentclass[12pt,preprint]{aastex631}
\usepackage{bbm}
\usepackage{mathrsfs}
\usepackage{amssymb,amsmath,float}
\usepackage{rotating}
\usepackage{color}
\usepackage{cancel}
\usepackage{ulem}
\usepackage[normalem]{ulem}

\newcommand\rh{\rho}

\newcommand\La{\Lambda}

\newcommand\Om{\Omega}

\newcommand\ie{\emph{i.e.}}
\newcommand\eg{\emph{e.g.}}

\newcommand\beq{\begin{equation}}
\newcommand\eeq{\end{equation}}
\newcommand\bea{\begin{eqnarray}}
\newcommand\eea{\end{eqnarray}}
\newcommand\bal{\begin{align}}
\newcommand\eal{\end{align}}

\newcommand\X{\times}
\newcommand\fr{\frac}

\renewcommand\d{\mathrm{d}}
\newcommand\e{\mathrm{e}}
\newcommand\h{\mathbbm{h}}
\renewcommand\i{\mathrm{i}}

\newcommand\R{\mathbb{R}}
\newcommand\C{\mathbb{C}}
\newcommand\Z{\mathbb{Z}}

\renewcommand\bal{\mbox{\boldmath$\alpha$}}

\begin{document}

\title{The interface of gravity and dark energy}

\author{Kristen Lackeos}{}
\affil{Max-Planck-Institut f{\"u}r Radioastronomie (MPIfR), Auf dem H{\"u}gel 69, 53121, Bonn, Germany\\}
\affil{NASA Postdoctoral Program Fellow, NASA Marshall Space Flight Center, Huntsville, AL 35812\\}
\author{Richard Lieu}{}
\affil{Department of Physics and Astronomy, University of Alabama, Huntsville, AL 35899\\}

\correspondingauthor{}
\email{klackeos@mpifr-bonn.mpg.de}

\begin{abstract}
At sufficiently large radii dark energy modifies the behavior of (a) 
bound orbits around a galaxy and (b) virialized gas in a cluster of galaxies. 
Dark energy also provides a natural cutoff to a cluster's dark matter halo.  
In (a) there exists a maximum circular orbit beyond which periodic motion is no longer possible, 
and orbital evolution near critical binding is analytically calculable using an adiabatic invariant integral.  
The finding implicates the study of wide galaxy pairs. In (b), dark energy necessitates the use of a generalized 
Virial Theorem to describe gas at the outskirts of a cluster.  When coupled to the baryonic escape condition, 
aided by dark energy, the results is a radius beyond which the continued establishment of a hydrostatic 
halo of thermalized baryons is untenable. This leads to a theoretically motivated virial radius. 
We use this theory to probe the structure of a cluster's baryonic halo and apply it to 
X-ray and weak-lensing data collected on cluster Abell 1835. We find that gas in its outskirts deviates 
significantly from hydrostatic equilibrium beginning at $\sim 1.3\ {\rm Mpc}$, the `inner' virial radius. 
We also define a model dependent dark matter halo cutoff radius to A1835. The dark matter cutoff 
gives an upper limit to the cluster's total mass of $\sim7\times 10^{15}M_{\odot}$. Moreover, it is possible 
to derive an `outer' hydrostatic equilibrium cutoff radius given a dark matter cutoff radius. 
A region of cluster gas transport and turbulence occurs between the inner and outer cutoff radii.
\end{abstract}

\section{Introduction}

The large scale structures we observe today are believed to have originated from seed density enhancements of the early universe that underwent linear gravitational growth until they `condensed' out of the Hubble flow and grew at an accelerated non-linear pace during the era of matter domination.  As dark energy started to dominate, the region where a structure `merges' with (or `interfaces') the cosmic substratum becomes one of particular interest and importance, for at least two reasons.


The first is the need to better understand the large scale dynamics of structures, in particular their mutual `action at a distance'  -- at what point does gravity finish and the repulsive force of dark energy take over?  If the universe has non-vanishing $\Lambda$, there will be a clear `zone of influence' of a structure, akin to the Debye shielding effect of electrostatics.  More precisely because gravity and dark energy act in opposition a `standoff' radius exists at which the two forces cancel each other, \ie~despite the long range effect of gravity (when it is the only force) a boundary must not only be in place but it must also be very sharp and well-defined.  In view of this, the evolution of galaxy pairs, specifically the so-called `wide pairs' at low redshifts (see \eg~Table 5 of \cite{nor98}, see also \cite{che93}), is especially affected by the cutoff and could even be used as a test bed of $\Lambda$,  see \cite{ben23}. 

The second reason, which is what mooted this paper, is about the largest bound structures, {\it viz.} clusters and groups of galaxies; in particular the distribution of baryons within them.  Although the total mass of one such system is dominated by the relatively cold and collisionless dark matter, the baryons are hot and virialized (or partially virialized) interacting particles. The shape of the density and temperature profile at the outskirts of a cluster or group is a topic of much current debate and investigation  (see \cite{gou23} and references therein).  Such studies are concerned with issues like the `gas fraction' (defined as the mass ratio $f_{\rm gas}$ of baryon to dark matter), as observations seem to indicate that this fraction is beneath the cosmic mean value in the interior of clusters, yet increasing monotonically towards the outskirts, pointing therefore to the phenomenon of baryonic escape, into regions outside the `grip' of the gravitational field of the structure, regions in which a hydrostatic halo of thermalized baryons cannot be established.

The structure of the paper is as follows. We begin in Section 2 by examining the evolution of bound orbits under the influence of the Hubble expansion in the dark energy dominated era. We consider three critical orbit scenarios for a test particle orbiting around a cluster or galaxy where the following condition holds: (a) the total energy per unit mass of a test particle is equivalent to the minimum effective potential and (b) $\lesssim$ the effective potential. Scenarios corresponding to circular and marginally bound orbits, respectively. Scenario (c) is the condition for the largest possible quasi-circular orbit defined for an unstable saddle point in the effective potential. We compare turnaround radii estimates from the literature, which can be defined for any epoch of the Universe, to the critical radius, $R_{\rm saddle}$, defined under scenario (c).
In Section 3 we expand on these three critical orbit scenarios by exploring changes to the total energy per unit mass and effective potential as the Universe evolves. Bound orbits in general are explored through the use of an adiabatic invariant, treating the dark energy component of the effective potential as a perturbation. From this we derive the perihelion and aphelion drift rates.
The virial theorem is enlisted in Section 4, to apply the findings of the previous two sections to baryons at the outskirts of galaxy clusters, and an application is made to cluster Abell 1835. Lastly, we conclude with a discussion and motivation to search for widely separated galaxy pairs and circumgalactic emission as a new means to experimentally test the dark energy paradigm.


\section{The effect of Hubble expansion on the evolution of bound orbits}

\subsection{Birkhoff Theorem}

We begin with a heuristic treatment of the motion of a small test particle of unit mass within some physical distance $R$ from the center of a large spherically symmetric mass concentration of mass excess $M$ in an otherwise homogeneous universe.  Since the environment experienced by the test mass is isotropic, Birkhoff Theorem can be applied, so that if one is in the weak gravity (large $R$) limit one may write down the energy equation  with $J$ being the angular momentum  as  \beq \dot R^2 + \fr{J^2}{R^2} = \fr{2GM}{R} + \fr{8\pi G}{3} \bar\rh R^2 + {\rm constant};  \label{energy} \eeq \ie~only the mass `underneath' the particle is relevant and the constant vanishes in the case of a flat universe (the value of this constant is unimportant to our work here).  Now (\ref{energy}) assumes that test particle is located at a sufficiently large $R$ for the central clump $M$ to be regarded as a point mass.  If the density at the position of the particle is above the mean density of the clump at that point, however, (\ref{energy}) will have to be modified to take account of the internal mass profile of the clump.

Turning to the mean background density, if the universe comprises non-relativistic matter of $\rh_m \sim 1/R^3$ and dark energy $\rh_\La =$ constant, one may write $\bar\rh = \bar\rh_m + \rh_\La$ and differentiate (\ref{energy}) w.r.t. the time $t$ whilst holding the angular momentum $J$ fixed by virtue of space isotropy, to obtain \beq \ddot R = -\fr{GM}{R^2} - \fr{4\pi G}{3} \bar\rh_m R + \fr{8\pi G}{3} \rh_\La R + \fr{J^2}{R^3}. \label{accel} \eeq  The relative sign difference between the matter and dark energy terms leads to two opposing forces acting upon the test particle, one attractive and the other repulsive.  At the critical radius where they balance, there is no radial acceleration of the particle, and bound orbits cannot exist beyond this radius.

To make contact with the mechanism underlying the apparent phenomenon of accelerated expansion, we must observe the effects of dark energy in our local universe. With a general relativistic treatment, we will see below that assuming a flat universe and using the Newtonian approximation are sufficient for our purposes. As first demonstrated by  \cite{mcc34}, `the relativistic and Newtonian theories are indistinguishable in their predictions of local phenomena'. They are the first to derive ($\ref{accel}$) using Newtonian physics only\footnote{Specifically, their equation of motion gives the path of a particle in a homogeneous universe~$\rho_m=\rho_m(t)$, where its motion satisfies the hydrodynamical equation of continuity.} and point out a repulsive force due to a cosmological constant, proportional to physical coordinate $R$. We extend this understanding to observable phenomena at the outskirts of large scale structures. 
\begin{center}
\end{center}

It is also possible to rewrite (\ref{accel}) as \beq \ddot R = -\fr{GM}{R^2} + \fr{J^2}{R^3} + H_0^2[\Om_\Lambda - \frac{1}{2} \Om_m (1+z)^3]R, \label{accele} \eeq where $H_0$ is the present Hubble constant and $\Om_m \approx0.3$ and $\Om_\La \approx0.7$ are the normalized densities.  Note that a proper derivation by \cite{ness04}, who used the Friedmann-Robertson-Walker metric as starting point, also yielded (\ref{accele}). 

Other metrics describing the geometry of a `massive structure in an expanding universe' have been shown to produce (\ref{accele}) in the Newtonian limit. \cite{bak01} demonstrates this using a metric originally derived by \cite{mcv33}. To the same end, \cite{nan121} (NLH1) derive a metric using the tetrad-based approach in general relativity, providing an exposition of the full general relativistic equations of motion, for all three curvature cases. In this way NLH1 point out errors in McVittie's open and closed universe solutions. And for a flat universe, they derive a simple coordinate transformation showing the metric they derive describes the same spacetime as McVittie's and reproduce (\ref{accele}), on the further assumption of low velocities.

\subsection{Bound orbits}

During earlier times $z \sim 0.67$ when the quantity inside the square parantheses of (\ref{accele}) is negative, there was another attractive force in additional to that of the central mass $M$, due to the cosmic substratum being matter dominated.   A bound orbit remains bound in those times.  In the recent dark energy era, however, the same quantity increases monotonically until it saturates to become a positive constant.  In this era, marginally bound orbits can become free.

One can gain a considerable amount of further insight on how orbits evolve by integrating (\ref{accele}) over a period in which the last term has not changed too much (such an interval ranges from many orbits to a segment of an orbit, depending on the size of the orbit and $M$) to obtain the energy equation \beq \fr{1}{2} \dot R^2 + V = E, \label{E} \eeq where $E=E(t)$ is the total energy per unit mass of the satellite (test) particle and \beq V(R,t) = \fr{J^2}{2R^2} - \fr{GM}{R} - \fr{1}{2} H_0^2 [\Om_\La - \fr{1}{2} \Om_m (1+z)^3]R^2 \\ = \fr{J^2}{2R^2} - \fr{GM}{R} -\gamma R^2 \label{Veff} \eeq is the effective potential.  The conservation of the `total energy' $E$ is of course only approximate.  Defining the symbols \beq q= \fr{\Om_m (1+z)^3}{2} - \Om_\La,~{\rm and}~\gamma = -\fr{1}{2} qH_0^2, \label{decel} \eeq with the former being the usual deceleration parameter and the latter a {\it positive} quantity during dark energy domination.  Note that $E$ changes at the rate \beq \dot E = -\dot\gamma R^2, \label{dotE} \eeq and this relation is {\it exact}.

The behavior of $V(R,t)$ is plotted in Figure 1 for cluster and galaxy-sized central bodies.  Several features are immediately apparent.  First, in the matter era {\it all} orbits are bound because the universe was collapsing upon itself.  Second, in the dark energy era the standard criterion of $E<0$ for bound orbits is no longer sufficient, as there is a maximum for $V$ at large $R$ which may be negative, and which in general is determined by the quantities $J$, $M$, $z$, and the two normalized densities $\Om_\La$ and $\Om_m$.  The radius $R$ of extremum $V$ is a solution of the quartic \beq GMR - J^2 - 2 \gamma R^4 = 0. \label{Vprime} \eeq For small angular momentum $J$ and large central mass excess $M$, or more precisely \beq J^2 \ll (GM)^{4/3} \gamma^{-1/3}, \label{compare} \eeq the minimum and maximum of $V$ occur at approximately \beq R_{\rm min} \approx\fr{J^2}{GM} + \fr{2\gamma J^8}{G^5 M^5};~R_{\rm max} \approx\left(\fr{GM}{2\gamma}\right)^{1/3}, \label{Rmaxmin} \eeq  with \beq V_{\rm min} =  -\fr{G^2 M^2}{2J^2} - \fr{\gamma J^2}{G^2 M^2},\\ V_{\rm max} = -3\left(\fr{\gamma G^2M^2}{4}\right)^{1/3}. \label{Vminmax} \eeq

\cite{che01} derives an $R_{\rm max}=(3M/(8\pi \rho_{\Lambda}))^{1/3}$, for a de Sitter universe, by equating a repulsive force, due to the total effective vacuum mass $M_{\Lambda}$ in a sphere of radius $R$, where $M_{\Lambda} \propto \rho_{\Lambda}R^3$, to the attractive force of a central mass $M$. \cite{che01} estimates in this way the radius for the Local Group where these two opposing forces equal. In the literature this radius defines the `zero gravity surface' of a large scale structure. Analytical and numerical solutions for orbits in a de Sitter universe are presented by \cite{em13} and \cite{che15}, respectively. We shall see below that the saddle point in the effective potential defines a cutoff for orbital boundedness in a $\Lambda$-dominated universe, which theoretically occurs sooner than $R_{\rm max}$, at $R_{\rm max}/ 2^{2/3}$.

\begin{figure*}
\centering
{\includegraphics[angle=0,width=6.5in]{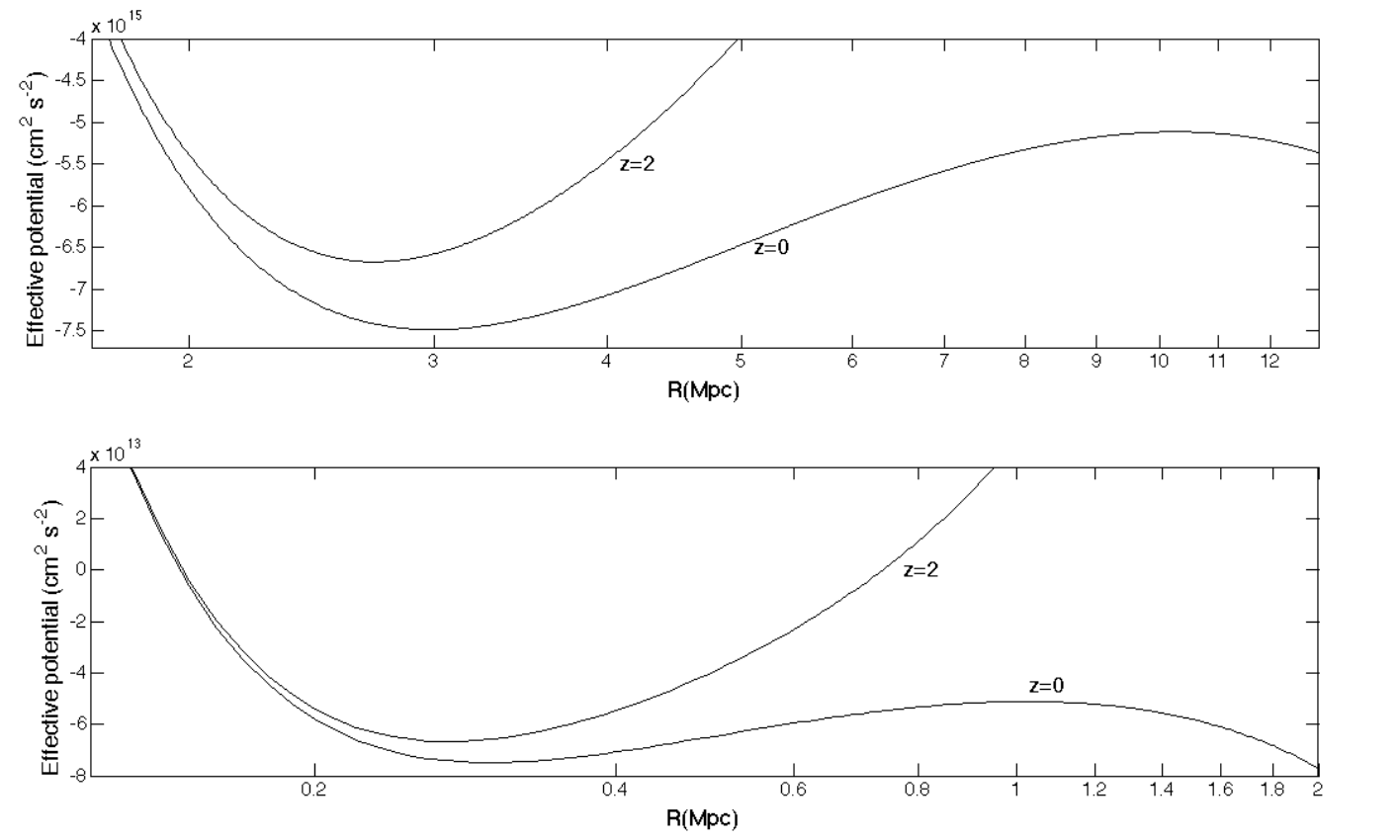}}
 \caption{Effective potential as given by (\ref{Veff}) plotted against the physical radius $R$ for a test mass orbiting a 10$^{15}~M_\odot$ cluster (top) and a 10$^{12}~M_\odot$ galaxy (bottom).  In each case, the epochs at which the potential is evaluated are $z=2$ (matter dominated) and $z=0$ (transition from matter to vacuum domination). It can be seen that the potential exhibits at maximum as $z \to 0$.  The angular momentum for each graph is set by the tangential physical velocity at $R=R_{\rm max}$, which is 350 km~s$^{-1}$ (top) and 35~km~s$^{-1}$ (bottom).}
 \label{1pdf}
\end{figure*}

We may now discuss critical orbits.  If (a) the energy $E=V_{\rm min}$, the orbit will be circular\footnote{{\it Strictly} circular orbits are forbidden for effective potential (\ref{Veff}), so in the following `circular' implies `quasi-circular'.}, with speed and period given respectively by  \beq v_\perp^2 = \fr{GM}{R_{\rm min}} - \fr{2\gamma J^8}{G^4 M^4 R_{\rm min}} \approx\fr{GM}{R_{\rm min}};~P= \fr{2\pi R_{\rm min}}{v_\perp}. \label{circ} \eeq  On the other hand, if (b) $E=V_{\rm max}$, the orbital period of this most elongated orbit (\ie~one with the aphelion at $R_{\rm max}$) will be \beq P = \fr{2\pi}{3\sqrt{6\gamma}} = \fr{2\pi}{3\sqrt{-3q}H_0} \label{period1} \eeq and is always comparable to the Hubble time.

Third, in the dark energy era there can be a scenario (c) under which the only bound orbit is circular.  This occurs when the inequality (\ref{compare}) is violated, and is instead replaced by the equation \beq J^2 = \fr{3}{8} \left(\fr{G^4M^4}{\gamma}\right)^{1/3}, \label{c} \eeq an equation that represents also the condition for the largest possible quasi-circular orbit, which is imminently unstable.\footnote{Relations similar to (\ref{c}) and (\ref{saddle}) are derived by \cite{bal06}, where they do the calculation for a de Sitter universe instead.} This occurs at the radius where $V$ reaches the saddle point $V'=V''=0$.  In fact, such a critical radius is \beq R_{\rm saddle} = \fr{4J^2}{3GM} = \fr{3GM}{4 v_\perp^2} = \fr{1}{2} \left(\fr{GM}{\gamma}\right)^{1/3}. \label{saddle} \eeq Thus this `special' orbit is unlike the usual circular orbits that satisfy the balance of centrifugal force and gravity, {\it viz.}~(\ref{circ}a), and the difference by the factor of $3/4$ is due to the role played by the repulsive force of the cosmic substratum.  Its period \beq P = \pi\sqrt{\fr{2}{3\gamma}} = \fr{2\pi}{\sqrt{-3q} H_0} \label{period2} \eeq is comparable to the Hubble time as well.

Relations (\ref{c}) and (\ref{saddle}) are derived by \cite{nan122} (NLH2), where they describe $R_{\rm saddle}$ as the radius of the largest {\it stable} circular orbit.  Rather, this outermost `nearly' circular orbit is imminently unstable, as explained in Section 3.1. NLH2 also derive (\ref{Rmaxmin}); however, our respective interpretations of $R_{\rm max}$ differ when it comes to the radial extent of hydrostatic equilibrium (H.E.), as shown in Section 4.

The instability of the outermost circular orbit provides us a testable hypothesis of a $\Lambda$-dominated universe. Dark energy disengages wide-pair galaxy systems, so one should find a statistical minority of interacting wide-pairs with projected separations $\sim$1 Mpc. The observed number of interacting wide-pairs should be no more than the statistical number calculated, assuming wide-pairs are coincident because the two incidentally passed within the other's `radius of attraction', \ie~by chance coincidence and not by tidal interaction. The interaction of galaxies in a pair may be determined from line-of-sight peculiar velocities, which is an observable in low redshift wide-pairs.

To gain a numerical feeling of the various quantities, one may calculate representative values of the critical radius, or radius of the `outermost bound orbit'.  We do it for galaxy and cluster sized central mass excess.  In the former, one finds \beq R= 0.7 \left(\fr{M}{10^{12} M_\odot}\right)^{1/3} \left(-\fr{q_0}{0.55}\right)^{-1/3} \left(\fr{h}{0.7}\right)^{-2/3}~{\rm Mpc},  \label{clusterR} \eeq and \beq v_\perp = 70~\left(\fr{M}{10^{12} M_\odot}\right)^{1/3} \left(-\fr{q_0}{0.55}\right)^{1/6} \left(\fr{h}{0.7}\right)^{1/3}~{\rm km}~{\rm s}^{-1}; \label{clusterv} \eeq while in the latter case one substitutes $M=10^{15} M_\odot$ to the above formulae to obtain $R \approx$ 7 Mpc and $v_\perp \approx$ 670~km~s$^{-1}$. NLH2 give comparable estimates for $R_{\rm saddle}$, for a galaxy and cluster.

A satellite orbiting around a cluster or galaxy cannot under any circumstance be gravitationally bound to either of them if the perihelion exceeds the value given by (\ref{clusterR}) for the appropriate $M$. For clusters the issue is academic in practice, because the timescale of a full orbital cycle exceeds the Hubble time for orbits of size at or near the scale in question (\ie~over such long periods (\ref{E}) no longer follows from (\ref{accele}).  Instead, the relevance of dark energy to clusters is its effect on diffuse virialized gas -- a topic to be discussed below.

There is another way to estimate the finite extent of large scale gravitationally bound structures, even for a matter-dominated universe.~Solving the equation of motion for the velocity of a test mass moving radially outward from the central mass clump gives the `surface of zero velocity' or `turnaround radius' to a clump. Turnaround may be defined for any epoch and depends on the initial energy $E(T)$ of the test mass, where $T$ is any particular time. Using the age of the Universe $T\approx1/H_0$ allows one to estimate an `interface' radius between the gravitational influence of the clump and the Hubble flow expansion at the present epoch. This is defined as a definite boundary to the structure today.~(\cite{lyn81}; \cite{san86})

Extending this `zero velocity' method to a dark energy universe, \cite{kar07} estimate the boundary radius and total mass of nearby groups and clusters of galaxies, using observed peculiar velocities of galaxies at the outskirts and Hubble flow data. With observational data on galaxies around the Local Group (LG), a boundary\footnote{\cite{kar09} show in Figure 1 of their paper radial velocities of galaxies in the LG and in the neighboring outskirts as a function distance with respect to the LG centroid. There appears to be a boundary to the LG at $\lesssim$ 1 Mpc.} to the LG is estimated to be $\lesssim$ 1 Mpc. Using this estimated radius, \cite{kar09} calculate the total mass of the LG to be $\sim 2\times 10^{12}M_{\odot}$. As expected, this boundary is numerically close to that predicted by (\ref{saddle}) above; their estimated mass for the LG gives $R_{\rm saddle}\approx 0.9$ Mpc.  In another recent study, \cite{nas11} used peculiar velocity data of more than 1000 galaxies at the Fornax-Eridanus complex outskirts (with radii $r\lesssim$ 30 Mpc) to  estimate the boundary radius and mass of the Fornax-Eridanus complex, obtaining $\lesssim$ 5 Mpc and ([1.30$-$3.93]$\times 10^{14}M_{\odot}$) respectively. Again using (\ref{saddle}), their mean estimated mass for the Fornax-Eridanus complex gives $R_{\rm saddle}\approx 4.5$ Mpc.


\section{The evolution of bound orbits}


The purpose of this work is to examine how bound orbits evolve in a dark energy universe (which necessarily means these orbit live in the present epoch $H=H_0$).  Therefore we will ignore such other dark energy induced phenomena as orbital precession (\cite{ker03}) which, though equally interesting, are not directly relevant.

We begin with the critical orbits, {\it viz.} the scenarios of (a) to (c) of the last section.

\subsection{Critical orbits}

In (a), if an orbit is initially circular, \ie~$R=R_{\rm min}$, by (\ref{dotE}) and (\ref{Veff}) $\dot E$ will to lowest order be constant and equal to $\partial V/\partial t|_R = -\dot\gamma R_{\rm min}^2$ as the satellite moves.  Thus, after one orbital period of $P=2\pi J^3/(GM)^2$  the position of $\dot R =0$ will be at the same radius as before, {\it except} that by then $V$ will no longer be minimized there because by (\ref{Rmaxmin}) $R_{\rm min}$ displaced by the amount $4\pi \dot\gamma J^{11}/(GM)^7$, \ie~the orbit has become slightly elliptical, with eccentricity $e= 4\pi J^9 \dot\gamma/(GM)^6$ and hence an average aphelion advancement rate of $\Delta R_{\rm aph}/P = 2\dot\gamma J^8/(GM)^5$, the latter being of order $H_0^3 R_{\rm min}^3/v_\perp^2$ which is very small (even for $R_{\rm min} = 1$~Mpc and $v_\perp = 300$~km~s$^{-1}$ the rate is only $\approx10$~km~s$^{-1}$). 

In (b), a marginally bound orbit with $E \lesssim V$ has, by (\ref{period1}), a period $P \approx1/H_0$, hence it is reasonable to assume that the satellite spends most of its time at or near the aphelion, \ie~$R \approx R_{\rm aph}$ is a constant, in which case $\dot E \approx-\dot\gamma R_{\rm aph}^2$, where $R_{\rm aph} \lesssim R_{\rm max}$ with the latter defined by (\ref{Rmaxmin}).  This means, in turn, that $|\dot E| < \dot\gamma R_{\rm max}^2 = 2^{-2/3} \dot\gamma (GM/\gamma)^{2/3} = -\dot V_{\rm max}$ where the last step is taken with the aid of (\ref{Vminmax}).  Thus it is evident that $E$ will eventually reach the level $V_{\rm max}$ from below.  At this point $E$ and $V_{\rm max}$ will go down at the same rate, since $R_{\rm aph} = R_{\rm max}$, while $R_{\rm max}$ continues to decrease by (\ref{Rmaxmin}) and the fact that $\dot\gamma$ is positive.  As time elapses further the satellite will find itself on the other side of the potential barrier, and its orbit has become unbound.

In (c), the only bound orbit possible is the largest circular orbit with $R=R_{\rm saddle}$.  From here onwards, $E$ and $V (R_{\rm saddle})$ go down with time in tandem while $R_{\rm saddle}$ also decreases, \ie~$R$ will soon become $R > R_{\rm saddle}$ and the satellite has escaped.

The following is a  discussion on the largest bound initially circular orbits around galaxies. This topic was first explored by NLH2, for orbits about clusters. Here we extend their discussion, for theoretical completeness only. The time scales involved are much longer than the age of the universe and involve unobservable future evolution. In Appendix I, we use our methodology to provide rough upper estimates of the largest initially circular orbits to remain bound indefinitely We do so for a test mass set in motion at various redshifts.~This is only an estimate, as the rate of decrease of $V_{\rm max}$ is greater than the orbital energy, so an orbit becomes unbound with less angular momentum than is required if the orbit were to remain exactly circular as the universe evolved. Namely, the test mass becomes effectively free at the time satisfying the condition $|E(t)|\leq |V_{\rm max}(t)|$, see Figure 6. For initial radii equal to critical this condition is met prior to the deceleration parameter saturating to $-0.7$. And for initial radii less than critical this condition is never satisfied, and the orbit remains bound indefinitely. 

We find the largest initially circular orbits, that are bound indefinitely, occur for an angular momentum less than the maximum at the saddle point angular momentum (\ref{c}) (the maximum occurs for $q_{\rm min}=-0.7$). That is, the maximum angular momentum allowed for {\it boundedness} is slightly less than that given by (\ref{c}) and is found by numerically evolving the orbit, giving the exact critical radius for any starting time. NLH2 find this critical radius for a test mass orbiting a cluster, beginning at the present epoch. In Appendix I we estimate, and calculate numerically, critical radii for test mass motion about a galaxy initiated at various epochs.

\subsection{Bound orbits in general, adiabatic invariant}

For an arbitrary bound orbit of $V_{\rm min} \ll E \ll V_{max}$, or more explicitly from (\ref{Vminmax}), \beq (G^2 M^2 \gamma)^{1/3} \ll |E| \ll \left(\fr{GM}{J}\right)^2, \label{bound} \eeq quantifying its evolution is a more arduous task because both the time dependence of $\dot\gamma$ and $R$ have to be taken into account (in particular, the change of $R$ with time as the satellite moves along such an `intermediate' orbit cannot be ignored) when integrating (\ref{dotE}) to produce $E(t)$.

Nevertheless, since the period of this class of orbits is $\ll 1/H_0$ it is possible to take advantage of the adiabatic invariant $$ I = \int_{R_{\rm per}}^{R_{\rm aph}} \sqrt{E-V(R)}~dR, $$ which may be evaluated by treating the $\gamma R^2$ term of the effective potential of (\ref{Veff}) as a perturbation (\cite{lan76}), {\it viz.} \beq I_R = \int_{R_{\rm per}}^{R_{\rm aph}} dR~\Bigg[\Big({E+\fr{GM}{R}-\fr{J^2}{2R^2}}\Big)^{1/2}\\ + \fr{\gamma R^2}{2} \Big(E+\fr{GM}{R}-\fr{J^2}{2R^2}\Big)^{-1/2}\Bigg]. \eeq The first term of the integrand leads to the standard result for the radial action variable $p_R$ of the Kepler problem, \ie~ \beq \int_{R_{\rm per}}^{R_{\rm aph}} dR~\left({E+\fr{GM}{R}-\fr{J^2}{2R^2}}\right)^{1/2}\\ = \pi\left( \fr{GM}{2\sqrt{-E}} - \fr{J}{\sqrt{2}}\right). \eeq   The second term may be handled by the substitution $$R= \fr{J^2}{GM}\fr{1}{1+e\cos\phi} $$  which converts the
integral to one of the form $\int d\phi/(1-e\cos\phi)^4$, resulting altogether in \beq I_R = \fr{\pi J}{\sqrt{2}}\left(\fr{1}{\sqrt{1-e^2}} -1\right)\\ + \fr{\pi\gamma J^7}{2\sqrt{2} G^4 M^4} (2+3e^2)(1-e^2)^{-7/2}, \label{IR} \eeq where \beq e = \left[1+2E\left(\fr{J}{GM}\right)^2\right]^{1/2} \label{e} \eeq is the lowest order contribution to the orbit eccentricity.

One can rewrite (\ref{IR}) in terms of yet another variable $y= 1/\sqrt{1-e^2} = GM/(J\sqrt{-2E})$, so that the statement of $I_R$ as an adiabatic invariant may be expressed as $$
\fr{\pi \gamma J^7 \dot y}{2\sqrt{2} G^4 M^4} (35y^6 - 15y^4) + \fr{\pi J}{\sqrt{2}} \dot y = - \fr{\pi \dot\gamma J^7}{2\sqrt{2} G^4 M^4} (5y^7 - 3y^5). $$  Under the criterion (\ref{bound}) for intermediate orbits $y \gg 1$, so that only the 3rd term on the left side and the 1st term on the right of the equation need to be kept, leading to the solution \beq E = - \fr{15^{1/3}}{2} (GM)^{2/3} (\gamma + \gamma_0)^{1/3}, \label{soln} \eeq where $\gamma_0 \gg \gamma$ is a constant.  This results in the rate of change of energy \beq \dot E = -\fr{5}{8}\left(\fr{GM}{E}\right)^2 \dot\gamma, \label{middle} \eeq which expectedly puts \beq \left(\fr{J^2}{2GM}\right)^2 \dot\gamma = \dot\gamma R_{\rm per}^2 \ll |\dot E| < \dot\gamma R_{\rm aph}^2 = \left(\fr{GM}{E}\right)^2 \dot\gamma, \label{inequality} \eeq where in arriving at (\ref{inequality}) use was made of (\ref{bound}).

It is now possible to calculate the speed and direction at which the aphelion and perihelion drift. Consider first the aphelion.  After one orbital period $P=\pi GM/\sqrt{-2E^3}$, the effective potential changes by $\Delta V = -\dot\gamma R_{\rm aph}^2 P$ while $E$ changes by $\Delta E=\dot E P$.  The displacement in the aphelion therefore given by the relation $V'(R_{\rm aph})\Delta R_{\rm aph} = \Delta E - \Delta V$.  Using (\ref{Veff}), (\ref{decel}), (\ref{middle}) and (\ref{inequality}) one then obtains the aphelion drift rate, to lowest order, \beq \fr{\Delta  R_{\rm aph}}{P} = \fr{3G^3 M^3}{8 E^4} \dot\gamma, \label{aphdrift} \eeq which indicates that the aphelion drifts {\it outwards} with the rate of order the product of the Hubble velocity at the distance $R_{\rm aph}$ ({\it viz.}~$H_0 R_{\rm aph}$) and the square of the ratio of this velocity to that of the orbit.  For $R_{\rm aph} \approx1$~Mpc and orbital velocity $\approx$ 300~km~s$^{-1}$, the rate is then of order 10$h^3$~km~s$^{-1}$.  A similar calculation for the perihelion leads to the rate \beq \fr{\Delta R_{\rm per}}{P} = \fr{5J^4}{32 GME^2}\dot\gamma, \label{perdrift} \eeq which is negligible compared with the aphelion by virtue of (\ref{bound}).

In summary, because both the perihelion and aphelion drift {\it outwards} with the latter at a much higher rate than the former, the orbit becomes more elliptical, \ie~intermediate orbits also evolve in the direction of becoming unbound.

\section{Virial Theorem in a dark energy universe, an application to cluster Abell 1835}

Apart from orbits, a gravitationally bound system also can also manifest itself as a spatially confined region of dark matter and thermalized baryons.  Notable examples are clusters and groups of galaxies that harbor a 10$^{6-8}$~K gas accounting for a good fraction of the baryonic mass of such structures.  The bulk of the X-ray and Sunyaev-Zeldovich measurements to date (\cite{vik06}, \cite{afs07}, \cite{arn07}, \cite{sun09}, and \cite{ett09}), with the notable exception of \cite{lan13} and \cite{sim11}, reveal that the mass ratio of baryons-to-total matter falls {\it below} the value predicted by the standard cosmological model, with a cluster's dark matter more centrally condensed than the baryonic matter, as the latter has a higher temperature.  Specifically, for a small radius $R$ the enclosed baryonic to dark matter mass ratio rises with $R$, but fails to reach the cosmic value of $f_{\rm gas} \approx1/7$ even at the limiting radius of X-ray sensitivity, which for a rich cluster is of order the so-called `virial radius' of $R_{\rm virial} = 2$~Mpc.

\subsection{Clusters bounded by purely gravitational effect}

We begin by addressing the question of what binds the baryons.  From numerical hydrodynamic simulations there emerged
a widely used spherically symmetric `universal' profile for the dark matter density distribution of clusters and groups, \ie~the matter that shapes the gravitational potential of these systems.  It is known as NFW profile, see Navarro et al (1995), (1996),
(1997),  and also the earlier work of \cite{dub91}, and has the form
\begin{equation}
\rho_m (R) = \frac{\rho_s R_s^3}{R(R +R_s)^2} \label{nfw}
\end{equation}
where $R$ is a physical radius,
$R_s$ is a constant scale radius, and $\rho_s$ is the central density parameter.  But since at the outer radii $R \gg R_s$ the density scales as $\rho_m (R) \sim 1/R^3$,
and there is no further change of functional form with increasing
distance, \ie~hierarchical structure formation codes do not seem
to reveal the surface radius of a clump,
the total integrated mass is divergent unless an upper limit
(or cutoff) radius $R$ is `manually' assigned via the `tapered NFW models' (\cite{spr98}, \cite{lok01}).

In the following we investigate the `lowest order'  effect of dark energy on baryonic gas at the cluster outskirts. We develop a theory that produces a cutoff radius to hydrostatic equilibrium $R_{\rm H.E.}=R_{\rm virial}$ and a separate cutoff to the dark matter enveloping the baryonic matter of a cluster. The latter cutoff radius is used to estimate the total mass of a cluster. We apply our theory to A1835, and with observations of the cluster we estimate its `true' virial radius and the extent of its dark matter halo and total mass.


\subsection{The outskirts of clusters in a dark energy universe}

For a cluster in Minkowski spacetime, up to a certain scale height, the `atmosphere' of baryons in its potential well could be approximated as an ideal gas in hydrostatic equilibrium with temperature profile $T(R)$ such that the mean kinetic energy $K$ equals half the mean potential energy $V$, or \beq 3kT = \fr{GM_cm}{R}, \label{virial} \eeq   where $m = \mu m_p \approx0.6m_p$, and $M_c=M_c(R)$ is the mass `underneath' a thin shell of radii $R$ and $R + dR$ . This equation of state ignores the external pressure. When we derive the Virial Theorem for structures in a dark energy universe below, the contribution to the mean kinetic energy from the external pressure is included.  

With a constant amount of underlying mass $M_c$, the mean kinetic energy per unit mass of the thermal gas is $\bar K = 3k_BT/(2m) = GM_c/(2R)$ for $R>R_{\rm virial}$, while at such radii the {\it extra} kinetic energy per unit mass, above the mean, required to enable an `average' particle to escape to infinity, the place where $V = 0$, is $K = GM_c/R$.  Thus a particle with kinetic energy satisfying the relation $K \geq 2\bar K$ can escape the gravitational pull of the cluster. To calculate the extent of the hydrostatic halo of baryons in the outskirts of a cluster, we must model a realistic scenario: we take account of the cosmology dependent background density, and we include mass and pressure profiles. The addition of these profiles allow us to include the effects of both the dark matter and gas at the outskirts of the cluster.

The escape condition is different in a $\Lambda$-dominated universe. A particle with kinetic energy in excess of its potential energy $V(R)$, in absolute value, {\it minus} $|V_{\rm max}|$ will escape: the condition to escape the cluster is $K \geq |V|-|V_{\rm max}|$. More generally, the criterion for escape is $K \geq \eta\bar K$, where $\eta$ becomes scale dependent: \beq \eta(R) = \fr{|V|-|V_{\rm max}|}{\bar K} \label{eta}. \eeq 


From (\ref{Veff}) we see that the potential per unit mass is now $V=V_1+V_2$ (the angular momentum term in (\ref{Veff}) plays no role for virialized systems). The dark energy contribution to $V$ is \beq V_2 = -\fr{1}{2}|q| H_0^2 R^2. \label{potent2} \eeq Owing to the presence of the $V_2$ term in the effective potential and to the external pressure from the outermost shell of baryons (immediately outside the hydrostatic halo)
the Virial Theorem is different from its usual version for gas trapped in a pure gravitational field.  Instead it is generalized to become \beq 2\bar K = \fr{3P(R)}{\rho_b(R)} - \bar V_1 +2 \bar V_2, \label{genvirial} \eeq which is the same as the equation of hydrostatic equilibrium in a dark energy universe. See Appendix II for derivation of (\ref{genvirial}) from H.E.. The kinetic energy per unit mass is $K$, where $\bar X$ denotes the average of $X$ over a timescale $\ll 1/H_0$. We assume $q=-|q_0|$ to find the largest $R$ satisfying (\ref{genvirial}) at the present epoch. 

We estimate the external pressure term in (\ref{genvirial}) with a temperature profile, where $P(R)/\rho_b(R)=k_BT(R)/(\mu m_p)$. We use temperature profile $T(R)$ from \cite{bur10}, with average temperature and fitting parameters from \cite{zha07} and \cite{ich13} respectively, for cluster A1835 \beq k_BT(R)=k_BT_{\rm avg}\times A\Big[1+B(R/R_{\rm 200})\Big]^{\beta}\ {\rm keV}.\label{tempprof} \eeq Where $k_BT_{\rm avg}\sim 7.67 \pm 0.21$ keV $A=1.74\pm 0.03$, $B=0.64\pm 0.10$, $\beta=-3.2\pm 0.4$ and $R_{200}\approx2.2$ Mpc (\cite{ich13}).

The total mass of the cluster inside radius $R$, (\ref{massbeyondvir}), is given by the full NFW profile, where we use parameters from the weak-lensing analysis of cluster A1835 (\cite{oka10}), taken in their current form from \cite{ich13}. `Virial mass' $M_{\rm \Delta}$ and halo concentration $c_{\Delta}=R_{\Delta}/R_s$ are 
\beq M_{\rm \Delta}=1.37^{+0.37}_{-0.29}\times 10^{15}M_{\odot}h^{-1}\label{mvir}\eeq 
\beq c_{\rm \Delta}=3.35^{+0.99}_{-0.79},\label{cvir}\eeq 
with Hubble parameter $h=0.7$. Up to $R_{\rm \Delta}$ we define the total cluster mass as follows, using (\ref{nfw}).\footnote{One may consult \cite{oka10}'s equations (11), (13), and (14) to see how we defined the central density parameter $\rho_s$ and scale radius $R_s$ in (\ref{massbeyondvir}), using only $M_{\rm \Delta}=M_{\rm virial}$, $c_{\Delta}=c_{\rm virial}$, $\Delta=\Delta_{\rm virial}=112$, and $z=0.253$.} 
\beq M(R)=4\pi \int^{R}_0\fr{\rho_sR_s^3}{r(R_s+r)^2}r^2dr. \label{Mr} \eeq  
For a cluster at low redshift, using the NFW density profile (\ref{nfw}), we define the mass profile by integrating (\ref{Mr}). 
This becomes \beq M(R)=4\pi \rho_sR_s^3\Bigg[{\rm Log}\Bigg(1+\fr{R}{R_s}\Bigg)-\fr{R}{R+R_s} \Bigg].\label{massbeyondvir} \eeq 
For $R=R_{\rm \Delta}$ in (\ref{Mr}) $M(R)$ equals $M_{\rm \Delta}$, of (\ref{mvir}), and the potential reduces to 
\beq V_1(R)=-G\fr{M_{\rm \Delta}}{R}. \label{potent1a}\eeq 
We use (\ref{potent1a}) to represent the inner region of the cluster. Otherwise we have the general NFW potential, 
\beq V_1(R) = -4\pi G\rho_sR_s^3\fr{{\rm Log}(1+\fr{R}{R_s})}{R}. \label{potent1b} \eeq 
The former potential satisfies Laplace's equation, and the latter (\ref{potent1b}) satisfies Poisson's equation, and the two, (\ref{potent1a}) and (\ref{potent1b}), will be used to define the criterion for escape,($\ref{eta}$). 

We use Sigmoid function \beq f=\fr{1}{{\rm exp}[\big(\fr{R}{{\rm Mpc}}\big)-\big(\fr{R_{\rm \Delta}}{{\rm Mpc}}\big)]+1}\label{sigmoid}\eeq to define $\eta(R)$, and require two conditions to be satisfied: $\eta(R)$ approaches 2 as $R\rightarrow 0$ and is continuous at $R=R_{\rm \Delta}=2.8$ Mpc. \beq \eta(R)=\eta(R\leq R_{\rm \Delta})\times f+\eta(R\geq R_{\rm \Delta})\times(1-f),\label{etafull} \eeq with \beq \eta(R\leq R_{\rm \Delta})=\eta(R,M_{\rm \Delta})\label{etaa}\eeq \beq \eta(R\geq R_{\rm \Delta})= \eta(R,M_{R\geq R_{\rm \Delta}}).\label{etab} \eeq We have from (\ref{eta}) and (\ref{genvirial}) that \beq \eta(R,M)=2\fr{|V|-|V_{\rm max}|}{-V_1+2V_2+3k_bT/(\mu m_p)}.\label{etaab}\eeq


Next we estimate the extent of the dark matter profile, which we assume cuts off at $R_{\rm max}$. The maximum of the cluster potential $V_{\rm max}$ occurs where the force of dark energy balances the gravitational attraction of the cluster's mass. We use (\ref{potent1b}) and (\ref{potent2}) to define the effective potential, $V_{\rm eff}(R)=V_1(R)+V_2(R)$ and use (\ref{massbeyondvir}) for the total mass. We find the maximum of the effective potential $V_{\rm max}$ is located at $R_{\rm max}\approx$ 17 Mpc. This gives an upper limit to the cluster's mass of $M_c(R_{\rm max})\lesssim 7\times 10^{15}M_{\odot}$. Using $R_{\rm max}$, and $M_c(R_{\rm max})$, we note $V_{\rm max}$ is constant in (\ref{etaab}).

The condition $\eta(R_{\rm H.E.})=1$ equates the kinetic energy of an average baryon, $K$, to the mean kinetic energy of a baryon, $\bar K$, given by the modified Virial Theorem (\ref{genvirial}). So baryons with kinetic energy equal to the mean at $R_{\rm H.E.}$ are not virialized and escape the cluster, hence $R_{\rm H.E.}=R_{\rm virial}$. 
Using the escape condition, $K\geq \eta \bar K$, we find $\eta=1$ at $R=R_{\rm H.E.}=1.3$ Mpc. See Figure 2.  Since baryons lack sufficient kinetic energy to remain virialized beyond $R_{\rm H.E.}$, they escape the gravitational well (with $K \geq |V|-|V_{\rm max}|$).

\begin{figure*}
\centering
{\includegraphics[angle=0,width=6.5in]{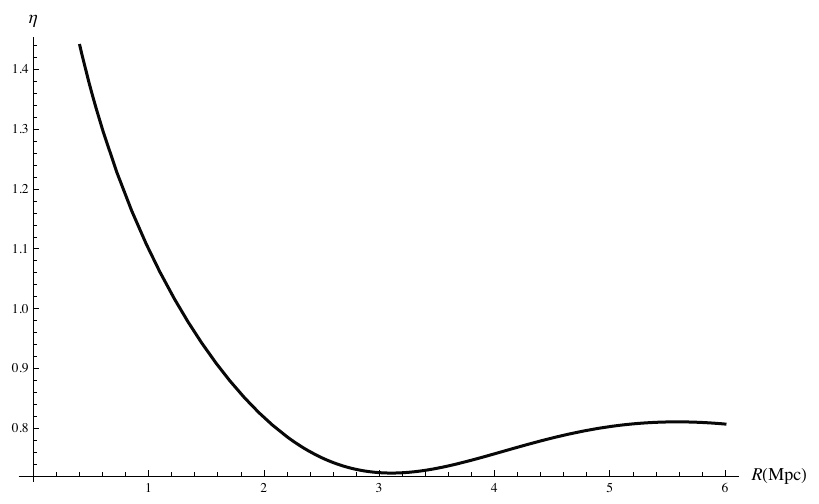}}
 \caption{A plot of Equation (\ref{etafull}), $\eta(R)$. The function approaches 1.9 as $R\rightarrow 0$, but is unphysical at zero, because $\eta$ is infinite at zero exactly. It is the contribution of $\eta(R\geq R_{\rm virial})$ that makes (\ref{etafull}) unphysical at small r, as it diverges at zero; however, $\eta(R\leq R_{\rm virial})$ (\ref{etaa}) converges to 2 as $R \rightarrow 0$ as expected.}
 \label{ichpdf}
\end{figure*}


We see in Figure (2) there is a complete cutoff at $R_{\rm H.E.}=1.3$ Mpc.
\cite{ich13}, \cite{kaw10}, and \cite{bau09} lend observational support to our theory, as each propose deviations from H.E. beginning from radii 1.4 Mpc, 1.7 Mpc, 1.3 Mpc, their respective values for $R_{500}$. In Figure (10) of \cite{ich13} they compare hydrostatic and weak lensing masses, one sees that a deviation from H.E. appears to begin as early as 1 Mpc even.

So with a cutoff to hydrostatic equilibrium at $R_{\rm H.E.}=1.3$ Mpc, we find the hydrostatic halo of baryons cuts off many times sooner than the dark matter profile at 24 Mpc.\footnote{NLH2 suggest the condition of H.E. in a dark energy persists out to a radius where the total pressure due to gas and dark matter equals zero, which they find occurs at $R_{\rm max}$. However, this may not be physically well-motivated at $R_{\rm max}$, because H.E. is inconsistent beyond $R_{\rm H.E.}$.}

To check the results of Figure 2 we evaluate the energy contribution of the external pressure term in (\ref{genvirial}) in a different way, this time using the pressure and electron number density profiles of (\cite{nag07}) and (\cite{vik06}) respectively, $P(R)/(2\mu m_pn_e(R))$. We assume $n_b=2n_e$ for a gas of thermalized baryons, with equal numbers of oppositely charged particles. We take from \cite{ade132} and \cite{bon12} best fitting X-ray pressure and number density profile parameters obtained from their respective observations of A1835. We reevaluate equations (\ref{massbeyondvir})-(\ref{etab}) with parameters from \cite{bon12} (Tables 3 and 4): the overdensity, total mass and halo concentration are $\Delta=100$, $ M_{\rm \Delta}\approx8.34\times 10^{14}M_{\odot}\label{mvircheck}$ (giving $c_{\rm \Delta}=5.8\label{cvircheck}$ and $R_{\rm \Delta}=2.2$ Mpc).  We alter (\ref{etaab}) to the following, substituting $P(R)/\rho_b(R)$ in place of $k_BT(R)/(\mu m_p)$ now. \beq \eta(R,M)=2\fr{|V|-|V_{\rm max}|}{-V_1+2V_2+3P/(\rho_b)}\label{etaabcheck}\eeq We find $\eta(R)=1$ at $R_{\rm H.E}$ = 2.0 Mpc, see Figure 3. From Figures 2 and 3, we conclude that dark energy affects a well-defined cutoff to H.E. between 1 and 2 Mpc. Using parameters from \cite{bon12} we find the maximum of the effective potential $V_{\rm max}$ is located at $R_{\rm max}\approx$ 17 Mpc. This gives an upper limit to the cluster's total mass of $M_c(R_{\rm max})\lesssim 2.2\times 10^{15}M_{\odot}$. 

\begin{figure*}
\centering
{\includegraphics[angle=0,width=6.5in]{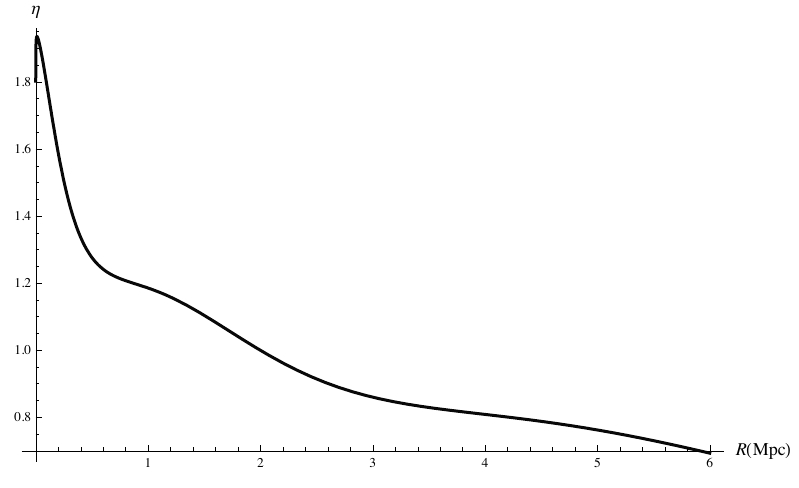}}
 \caption{A plot of Equation (\ref{etafull}), $\eta(R)$. Radius $R_{\rm H.E.}$ moves out from 1.3 Mpc to 2.0 Mpc when we use $P/\rho_b$ instead of $k_BT/(\mu m_p)$. See (\ref{etaabcheck}) in contrast to (\ref{etaab}).}
 \label{bon1pdf}
\end{figure*}

The hydrostatic mass of A1835 calculated by \cite{bon12} may be an underestimate of the total mass at larger radii, see \cite{ich13} Figure 10 comparing hydrostatic and weak-lensing masses. \cite{lan13} suggest the hydrostatic mass underestimates the true mass also. To account for this possibility we use equations (\ref{massbeyondvir})-(\ref{etab}) with parameters (\ref{mvir}) and (\ref{cvir}) (from the weak-lensing analysis of \cite{oka10}), as in our first calculation, but, for our check, continue to use $P(R)/(2\mu m_pn_e(R))$ in place of $k_BT/(\mu m_p)$. When using the larger lensing mass of \cite{oka10}, we find $\eta(R)=1$ at $R_{\rm H.E}$ = 2.1 Mpc, with the same dark matter cutoff and total cluster mass as in our first calculation of $R_{\rm H.E.}$. 

From Sunyaev-Zel'dovich observations of the Coma cluster by the Planck satellite (\cite{ade13}), we note an `atmosphere' of hot gas continues outside $R_{\rm H.E.}$; the hot gas atmosphere of the Coma cluster is found to extend out to 4~Mpc, at least. Since a cluster's dark matter cuts off at $R_{\rm max}$, we recalculate $\eta(R)$ for A1835. We ignore its inner structure and represent it as a point mass. This leads to an `outer' virial radius.

\subsection{{Constraining dark energy by observation of large scale structures}}

\begin{center}
\end{center}

In view of the aforementioned considerations, dark energy may be scrutinized by observing wide galaxy pairs and the thermal baryon envelope in clusters of galaxies.

One specific example of the first method is the galaxy pair of the Milky Way and the Andromeda galaxy. This is a wide pair because the Keplerian orbital period $\approx 17$ Gyr, which exceeds the age of the Universe.  Thus dark energy plays a crucial role in the evolution of the pair.  An upper limit on $\Omega_{\Lambda}$ based on current mass estimates of both galaxies is provided by \cite{ben23}.

On the second method, 
 baryonic gas crossing the cutoff surface defined by $R_{\rm H.E.}$ may `escape' the cluster to radii greater than $R_{\rm max}$, by way of a cluster wind for example, but indefinite escape is not a guarantee. A halo of thermalized baryons may exist farther out too, but inside $R_{\rm max}$ still, with a terminating surface at an `outer' virial radius defined analytically next.  One way to see this is to recalculate $\eta(R)$ assuming the upper total mass limit of the cluster given by (\ref{Mr}) evaluated at $R_{\rm max}$ (for A1835 $R_{\rm max}\approx 24$ Mpc). In Figure 4 we show that hydrostatic equilibrium remains an option at larger radii for baryons transported outward by cluster winds and in general for baryons unsupported by H.E. at the inner cutoff radii derived above.

\begin{figure*}
\centering
{\includegraphics[angle=0,width=6.5in]{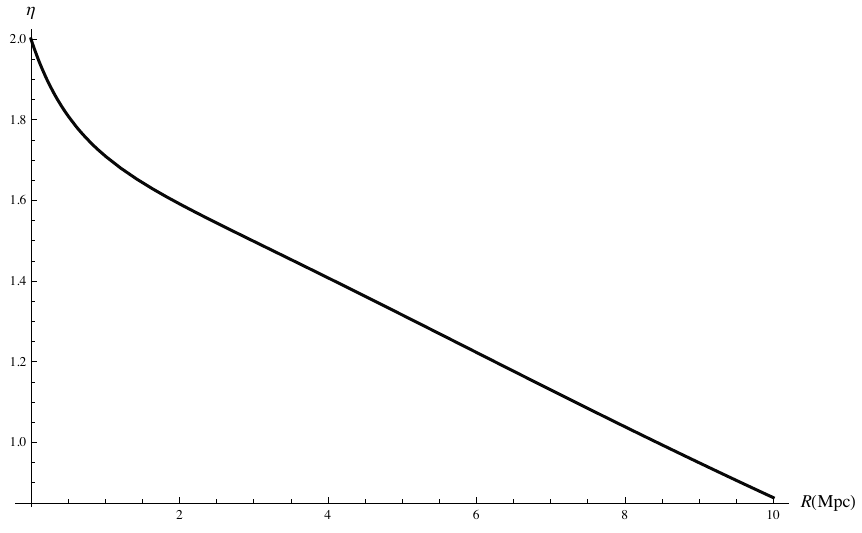}}
\caption{A plot of Equation (\ref{etafull}), $\eta(R)$. Using the dark matter cutoff of $R_{\rm max}$ the cluster is modeled as a `point mass',  and hydrostatic equilibrium persists out to $\sim$ 8.4 Mpc, which is comparable to our analytically derived result (\ref{Re}).}
 \label{eta_vmaxpdf}
\end{figure*}


Applying the condition $\eta$~=~1 to A1835, the radius satisfying (\ref{eta}) ($\bar K=|V|-|V_{\rm max}|$) and (\ref{genvirial}) is calculated analytically and compared with $R_{\rm max}$. The maximum of the effective potential $V_{\rm max}$ is still given by (\ref{Vminmax}b) for constant $M$. In solving (\ref{escR}) we shall find that $R_{\rm H.E.} < R_{\rm max}$.  \beq \fr{GM_{\rm max}}{2R_{\rm H.E.}} + |q| H_0^2 R_{\rm H.E.}^2 - \frac{3}{2} |q|^{1/3} (GM_{\rm max}H_0)^{2/3} = 0. \label{escR} \eeq   
By rewriting in terms of the dimensionless variable $x=RH_0/c$ (\ref{escR}) becomes $$ x^3 - ux+v = 0, $$ where $u = 3|q|^{-2/3} (GM_{\rm max} H_0)^{2/3}/(2c^2)$ and $ v=GM_{\rm max} H_0/(2|q|c^3)$ are also dimensionless.  In a cluster environment where $GM_{\rm max}/R \ll c^2$ and $R\ll c/H_0$, it can readily be shown that $u \gg v$.  Moreover, in a $\Omega_\Lambda = 0.7$ and $\Omega_m = 0.3$ cosmology where (\ref{decel}a) gives $q = -0.41$ for $z=0.253$, it can further be shown that $u^3 \gg v^2$.  Under these conditions, the $x^3$ term of the cubic is of little importance, \ie~$x=v/u$ or \beq R_{\rm H.E.}\\ = 8.02 \left(\fr{M_{\rm max}}{6.76\times10^{15}~M_\odot}\right)^{1/3}\left(\fr{|q|}{0.41}\right)^{-1/3} \left(\fr{h}{0.7}\right)^{-2/3}~{\rm Mpc}. \label{Re} \eeq  This is to be compared with \beq R_{\rm max} = \left(\fr{GM_{\rm max}}{|q|H_0^2}\right)^{1/3} \\ = 24.06~\left(\fr{M_{\rm max}}{6.76 \times 10^{15}~M_\odot}\right)^{1/3}\left(\fr{|q|}{0.41}\right)^{-1/3} \left(\fr{h}{0.7}\right)^{-2/3}~{\rm Mpc}.  \label{Rmax} \eeq Thus we see that $R_{\rm H.E.} \approx R_{\rm max}/3$ in general. Using $R_{\rm max}=17$ Mpc gives an $R_{\rm H.E.}=5.5$ Mpc.

For A1835 the latter is close to the $R_{500}$ radius obtained from Suzaku X-ray observations of the intracluster medium of cluster A1835 (\cite{ich13}) and the circumgalactic medium of the Illustris-TNG simulations (\cite{gou23}).   
 It is now apparent that in a dark energy universe the atmosphere of thermal gas bounded by the gravitational field of a central mass excess has a clear boundary.


\section{Discussion and conclusion}

The manner in which dark energy (a) disengages binary systems, and (b) places a hard upper limit on the extent of any baryonic halo in a cluster, is demonstrated in some detail. In (a), it is shown that closed orbits can no longer exist, and all orbits evolve to become less bound.  Moreover, the path of this evolution can be determined by means of an adiabatic invariant in most cases.  The phenomenon provides another way of testing the dark energy interpretation, via the observation of wide ($\sim$ 1 Mpc) galaxy pairs at low redshifts, as their line-of-sight velocities should indicate that they are at best only marginally bound.
In (b), we find that the thermal escape velocity of the `last scattering surface' of the atmosphere of hot baryons in clusters and groups of galaxies is lowered by dark energy, which causes the potential function to exhibit a sub-zero maximum.  The radius $R_{\rm H.E.}$ that marks this limit.   The results of this paper can be tested by observation of wide galaxy pairs and circumgalactic emission (see previous section).

\appendix
\renewcommand{\thefigure}{A\arabic{figure}}
\setcounter{figure}{0}
\section[]{Estimating the largest initially circular orbits to remain bound indefinitely}

To estimate the largest circular orbit that remains bound for all time, we equate the angular momentum of a test particle in an initially circular orbit, set in motion at any redshift, to the maximum angular momentum found at the saddle point, \ie~at $q_{\rm min}=-0.7$. If the test mass lacks sufficient energy to escape by the time $q$ saturates to $q_{\rm min}$, the orbit remains bound indefinitely. The critical radius for a particular initial $q$ is approximately given by \beq GMR_c(t)+q(t)H_0^2R^4_c(t)=\fr{3}{4} \left(-\fr{G^4M^4}{4 q_{\rm min}H_0^2}\right)^{1/3}. \eeq \label{rc} 
A test mass on an initially circular orbit with initial radius $< R_c$ will remain bound indefinitely. For an initially circular orbit around a $10^{12}M_{\odot}$ galaxy, starting in the present epoch, $q(t_0)=-0.55$ and $H_0=70\ \rm km\ s^{-1} Mpc^{-1}$, (39) gives $R_c(t_0) \approx0.58$ Mpc, and by evolving the orbit numerically we find $R_c(t_0)\approx0.57$ Mpc. Moreover, with (39) we find valid critical radii estimates when initiating the evolution from any epoch.

\subsection{Exact critical radii from numerical experiment}

\begin{figure*}
\centering
{\includegraphics[angle=0,width=6.5in]{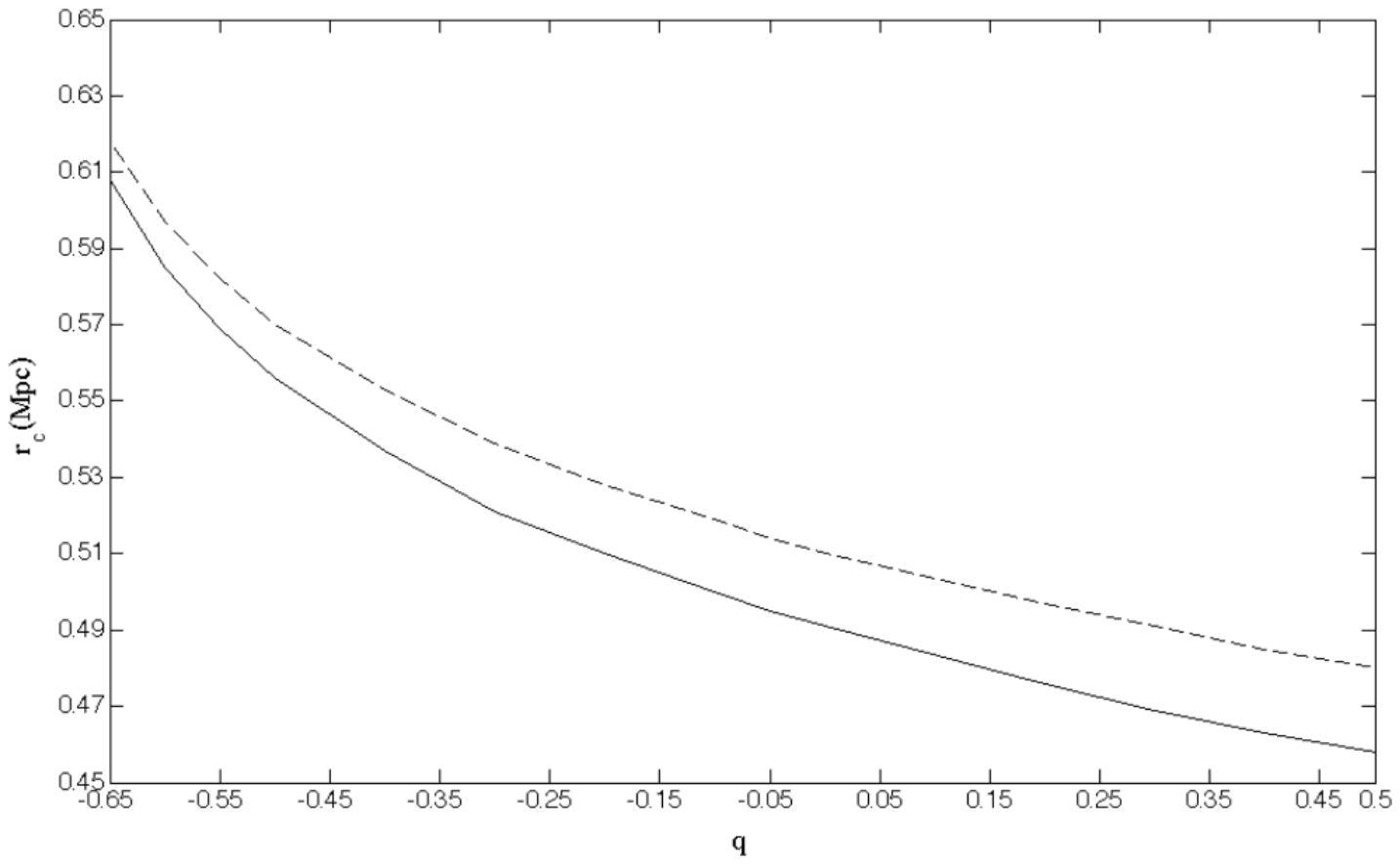}}
\caption{For various initial $q$'s, corresponding to different starting times, the largest initially circular orbit to remain bound after $q$ saturates to $q_{\rm min}$ is estimated and calculated numerically. The dashed curve gives critical radii estimated using (39), and the solid curve is from numerical evolution. \newline By increasing $R_c$ by one order of magnitude, one obtains an analogous plot for a cluster sized central mass, $M=10^{15}M_{\odot}$. The same applies to $r$ in Figures (\ref{fig5}), (\ref{fig6}) and (\ref{fig7}) below.}
 \label{fig5}
\end{figure*}

\begin{figure*}
\centering
{\includegraphics[angle=0,width=6.5in]{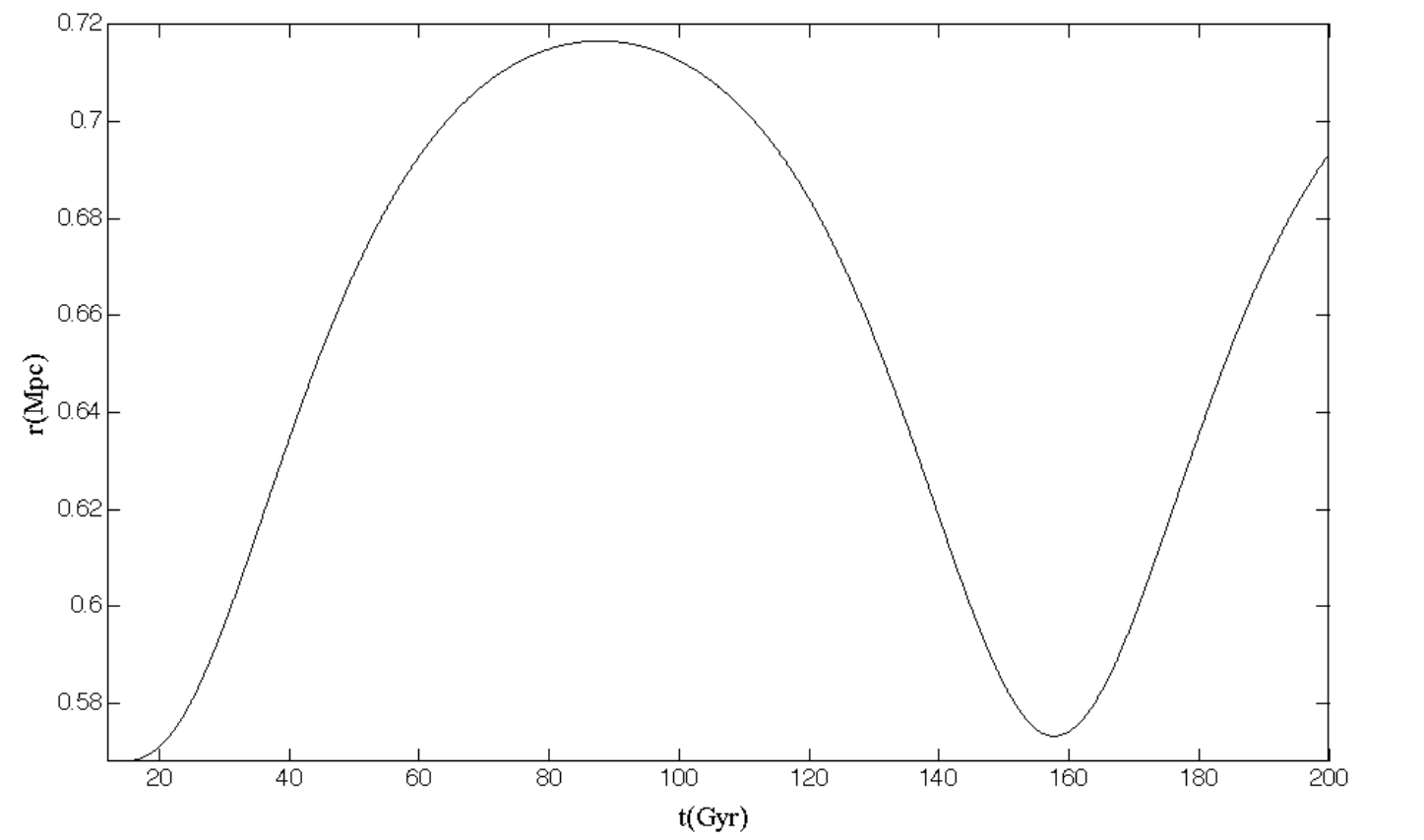}}
\caption{Starting the numerical evolution at $R_i=0.568$ Mpc, for $q_i=q_0$, the test mass oscillates about $R=0.642$ Mpc indefinitely.}
 \label{fig6}
\end{figure*}

\begin{figure*}
\centering
{\includegraphics[angle=0,width=6.5in]{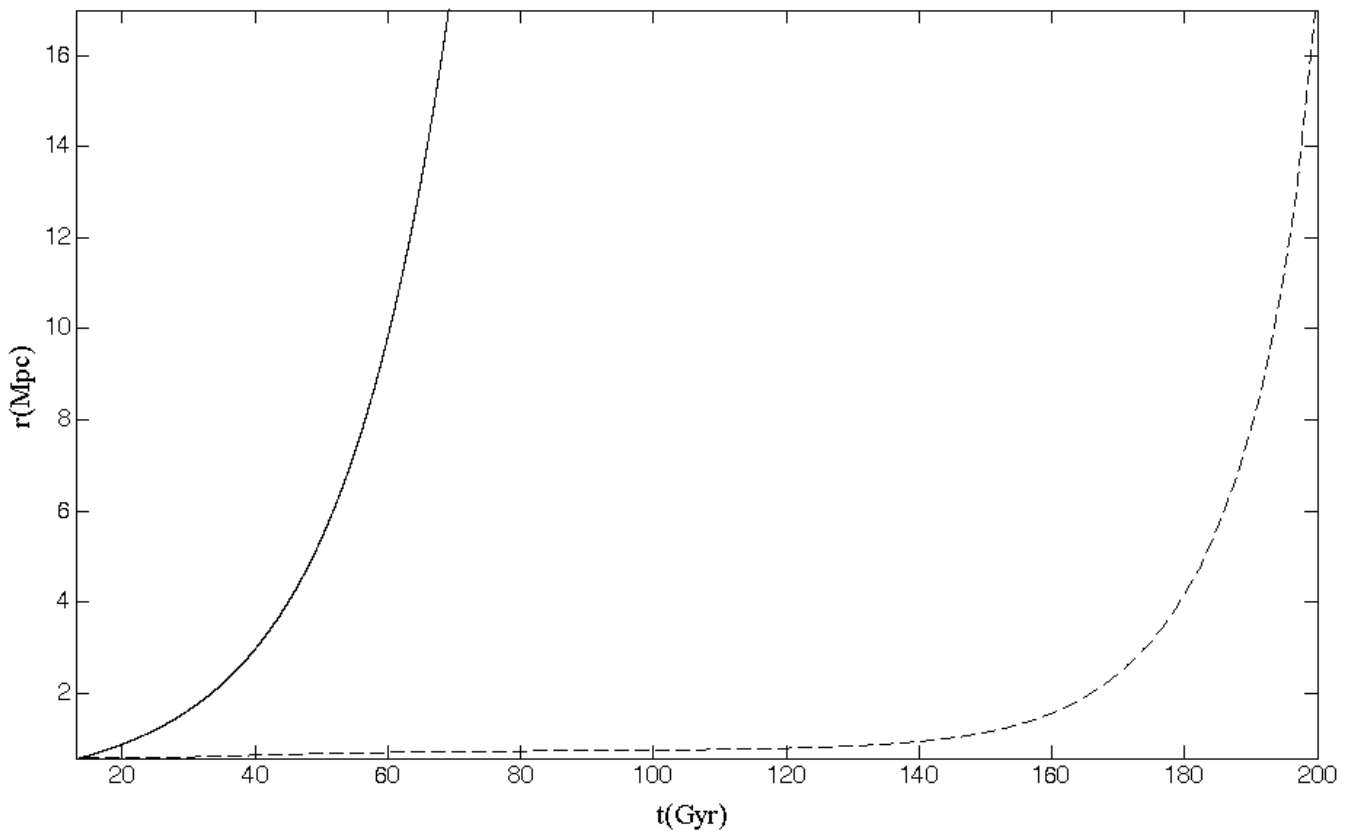}}
\caption{Starting the numerical evolution at $R_i=R_c=0.569$ Mpc, for $q_i=q_0$, the test mass escapes the potential well. As pointed out by NLH2, its motion, dashed curve, lags behind the Hubble expansion, solid curve, by a significant margin.}
 \label{fig7}
\end{figure*}

\begin{figure*}
\centering
{\includegraphics[angle=0,width=6.5in]{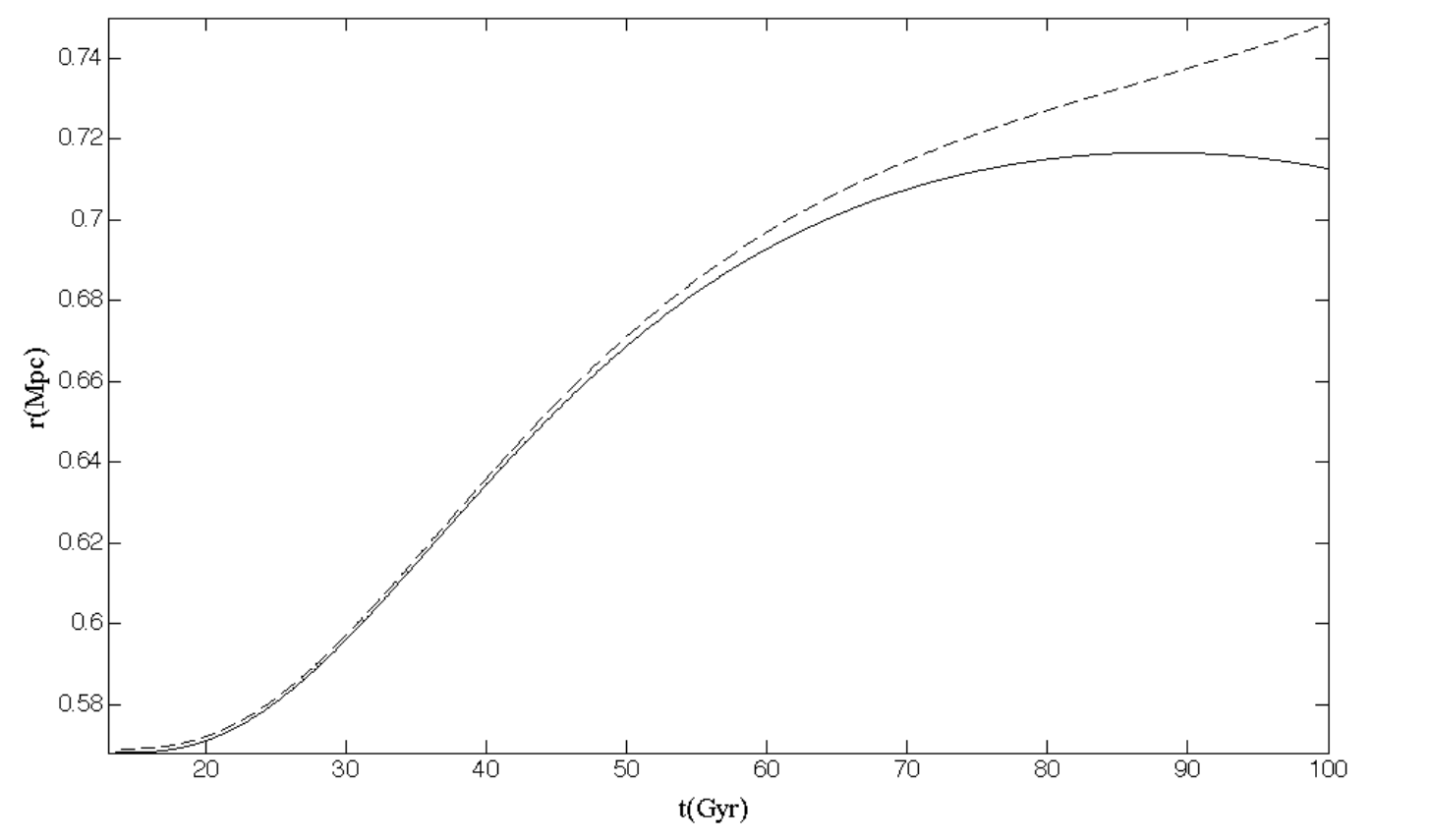}}
\caption{Figures (\ref{fig5}) and (\ref{fig6}) together, comparing test mass evolution for the first 100 Gyr.}
 \label{fig8}
\end{figure*}

\begin{figure*}
\centering
{\includegraphics[angle=0,width=6.5in]{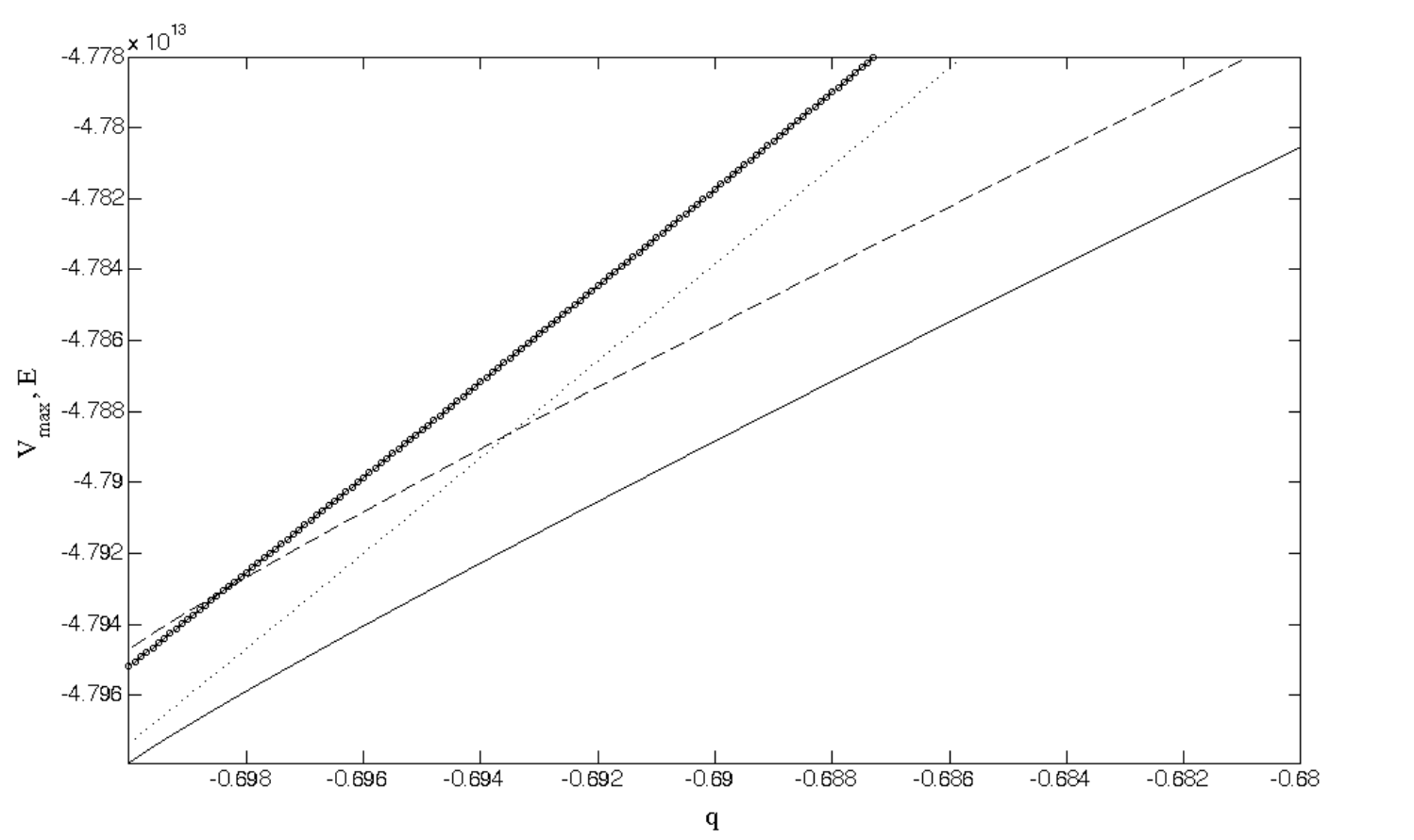}}
\caption{Continuing the scenario presented in Figures (\ref{fig5})-(\ref{fig7}), with the evolution beginning at $q_i=-0.55$, the total energy per unit mass of the orbit as a function of $q$ is calculated by solving (10) numerically, and is compared to the exact maximum of the effective potential per unit mass. The curve with open circles is $V_{\rm max}$, with $R_i=R_c=0.569$ Mpc, and the dotted curve is $V_{\rm max}$, with $R_i=0.568$ Mpc. The dashed curve is the energy of the test mass, with $R_i=0.569$ Mpc, and the solid curve is the energy of the test mass, with $R_i=0.568$ Mpc. For both starting radii the test mass is initially bound, having energy `below' the respective $V_{\rm max}$. When the test mass is set in motion at the critical radius, its energy goes `above' $V_{\rm max}$, and so is able to escape the potential well, \ie~prior to $q$ saturating to $q_{\rm min}$, though, as shown in Figure (\ref{fig6}), its subsequent path lags behind the universal expansion. When starting from smaller radii its energy is below $V_{\rm max}$ indefinitely.}
 \label{fig9}
\end{figure*}

We check the validity of our estimated critical radii for a galaxy, by evolving (6) numerically, as NLH2 have done for a test mass about cluster. We use the universal expansion factor for a flat, $\Lambda$CDM universe. From \cite{pee93}, we use expansion factor \beq a(t)=\Bigg[\fr{\Omega_{m}}{1-\Omega_{m}}{\rm sinh^2}\Bigg(\fr{3}{2}(1-\Omega_{m})^{1/2}H_0t\Bigg)\Bigg]^{1/3}. \eeq We use equation of motion (\ref{accele}), where the starting time is found for various $q$'s using \beq q(t)=\fr{\Omega_ma(t)^{-3}}{2}-\Omega_{\Lambda}.\eeq We take a galaxy sized central mass $M=10^{12}M_{\odot}$, $\Omega_{\Lambda}=0.7$, $\Omega_{m}=0.3$, and constant (specific) angular momentum $J=\Big(GMR_i+q_iH_0^2R_i^4\Big)^{1/2}$, \ie~for a test mass in an initially circular orbit. We start the orbit with an initial radius near the appropriate critical radius estimated from (39). By numerical experiment (trial and error) we find the actual $R_c=R_i$ for which any initially circular orbit with radius greater than or equal to this will become free, prior to the saturation of $q$ to $q_{\rm min}$. In Figure (\ref{fig5}) we compare estimated and numerical critical radii for various initial $q$'s. In Figures (\ref{fig6}), Figure (\ref{fig7}) and Figure (\ref{fig8}) we illustrate behavior near the critical initial radius for $q=q_0=-0.55$. And in Figure (\ref{fig9}) we compare the evolution of the orbital energy to the evolving effective potential maximum, for an initial radius at critical and less than critical.

\renewcommand{\thefigure}{B\arabic{figure}}
\setcounter{figure}{0}
\section[]{Generalized Virial Theorem from hydrostatic equilibrium}
\setcounter{equation}{0} 
\renewcommand{\theequation}{B\arabic{equation}}

\beq \fr{dP}{dR}=-\rho_b(R)\fr{GM_{\rm Total}(R)}{R^2} \label{HE} \eeq
\noindent
Multiply through by the volume and integrate both sides, where we use $dM_b=4\pi R^2 \rho_b(R)dR$ to affect a change of integration variables on the RHS.



$$ \int_{0}^{P(R)}VdP=-\fr{1}{3} \int_0^{M_b(R)} \fr{GM_{\rm Total}(R')}{R'}dM_b'=\fr{1}{3}\Phi(R) $$

\noindent
Integrate the LHS by parts.
$$ \left. PV \right|_0^R  - \int_{0}^{V(R)}PdV =\fr{1}{3}\Phi(R) $$


$$ P(R)V(R)  - \int_{0}^{M_b(R)}\fr{k_BT}{\mu m_p}dM'_b =\fr{1}{3}\Phi(R) $$


\noindent
Substituting the average energy per baryon $\bar T=3/2k_BT$, we have the familiar Virial Theorem with an external pressure term:

$$ 3\fr{P(R)}{n_b}  - 2\bar T =-\fr{Gm_bM_{\rm Total}(R)}{R}. $$

Adding the force of dark energy per unit volume $\rho_b(R)|q_0|H_0^2R$ to the RHS of (\ref{HE}) we find the generalized Virial Theorem: 

$$ 3\fr{P(R)}{n_b}  - 2\bar T =-\fr{Gm_bM_{\rm Total}(R)}{R} +m_b|q_0| H_0^2R^2. $$

\noindent
Dividing through by the mass of a baryon $m_b$ we arrive to (\ref{genvirial}) of the text, where the average energy per unit baryon mass is $\bar K=\bar T/m_b$.

$$    2\bar K =-V_1 + 2V_2+3\fr{P(R)}{\rho_b} $$

\end{document}